\documentclass[11pt]{article}
\newlength{\vshift}
\newlength{\hshift}
\usepackage{amssymb}
\usepackage{amsmath}
\usepackage{amsthm}
\usepackage{amsopn}

\def\La{\Lambda}

\def\th{\theta}

\def\de{\delta}

\def\p{\partial}

\def\lb{\lbrack}
\def\rb{\rbrack}

\def\beq{\begin{equation}}
\def\eeq{\end{equation}}
\def\bea{\begin{eqnarray}}
\def\eea{\end{eqnarray}}
\begin{document}

\vspace{4em}
\begin{center}

{\Large{\bf Bounds on the Parameter of Noncommutativity from
Supernova SN1987A }}

\vskip 3em

{{\bf M. Haghighat \footnote{e-mail: mansour@cc.iut.ac.ir }}}

\vskip 1em

Department of Physics Isfahan University of Technology, Isfahan,
84156-83111, Iran
 \end{center}
\begin{abstract}
We consider supernova SN1987A to find bounds on parameter of
noncommutativity, $\th_{\mu\nu}$. Right handed neutrino in the
noncommutative standard model, NCSM, can directly couple to the
photon and the Z-gauge boson. Therefore the observed flux of
neutrinos from SN1987A can constrain the strength of the new
couplings in the NCSM. We obtain two bounds on the NC-parameter,
$\Lambda_{NC}=1/{\sqrt |\th|} $, with respect to escaping or
trapping of the right handed neutrinos inside the supernova which
 are $\Lambda_{NC}\gtrsim 3.7 TeV$ or $\Lambda_{NC}\lesssim 1  TeV$,
 respectively.  The excluded region $1 TeV\lesssim\Lambda_{NC}\lesssim
 3.7 TeV$ for the NC-parameter is obtained for the first time.  In fact $\Lambda_{NC}\lesssim 1
 TeV$ is consistent with the existing bounds on $\Lambda_{NC}$ and raises our hopes to find the NC-effects in the LHC or even in the LEP.
 Meanwhile $\Lambda_{NC}\gtrsim 3.7 TeV$ is more stringent than the
 other bounds obtained from LEP and LHC considerations.
 Furthermore, since NC-calculations are perturbative and are correct only up to
  the energy scale of the NC-system $\lesssim\Lambda_{NC} $, therefore
 these bounds which are obtained from the energy scale considerably less than the energy scale of LEP and LHC are more
 reliable.
\end{abstract}

Noncommutative spaces and their phenomenological aspects have been
extensively considered by many authors for a decade
\cite{sw}-\cite{positron}. The natural scale for the
noncommutativity effects seems to be the Planck scale which is far
from the terrestrial experiment to reach. However, there is no
theoretical constraint for the energy scale and one may think this
scale might be broken by some mechanism down to the lower energy
range and thus be testable by, for example, the Large Hadron
Collider (LHC) that is recently started to run.  There are two
approaches to construct the standard model in the noncommutative
space.  Nevertheless, for the practical purpose, both versions are
based on the perturbative expansion in terms of the parameter of
noncommutativity, $\th_{\mu\nu}$. Actually the region where the
$\th$-expansion is well defined is restricted to $\th_{\mu\nu}p^\mu
q^\nu\lesssim 1$ or ${\frac{E}{\Lambda_{NC}}}\lesssim 1$ where
$E\thicksim\sqrt s$ is the energy of the system.  Therefore one can
not a priori assume $\Lambda_{NC} \gtrsim \sqrt s$ and then compare
the new interactions which arise from the perturbative expansion to
put bounds on the $\Lambda_{NC}$.  For instance if
$\Lambda_{NC}\backsim 200 GeV$, the comparison of an obtained
amplitude in the NC-space up to the first order of the NC-parameter
with the experimental data of the corresponding scattering in $\sqrt
s \thicksim 1 TeV$ in order to find a bound on $\Lambda_{NC}$ is
obviously incorrect.  The existing bounds greater than $\thicksim 1
TeV$ are usually obtained by assuming $\Lambda_{NC}\gtrsim 1 TeV $
for the experiments with $\sqrt s \gtrsim 1 TeV$.  Therefore
Low-energy experiments tend to provide more reliable limits on
$\th_{\mu\nu}$.  There is no doubt that the parameter of
noncommutativity is greater than the atomic energy scale.
Therefore, one can begin with the atomic constraints such as the
Lamb shift in the hydrogen atom \cite{hydrogen}, the positronium
hyperfine splitting or the transitions in the Helium atom
\cite{positron} to find $\Lambda_{NC}\gtrsim 6 GeV, 1 MeV, 30 GeV$,
respectively. Since from the atomic energy scale $\Lambda_{NC}$ is
constrained to be greater than $30 GeV$ we can safely apply the
perturbative expansion to put bounds on the $\Lambda_{NC}$ in those
processes with the energies up to a few $GeV$.  On the other hand,
astrophysical limits can usually provide much more restrictive
bounds on strength of new couplings and existence of new particles
beyond the standard model than any laboratory experiment could
achieve. For example the neutrino burst duration of supernova, SN,
1987A is considered by many authors to find limits on photons
escaping into extra dimensions \cite{extra}, unparticles
\cite{unparticle}, light dark matter particles \cite{dark}, neutrino
oscillations \cite{neutrino}, Kaluza-Klein graviton production
\cite{klein}, sterile-neutrino production \cite{sterile}, axions
\cite{axion}, magnetic moment of the neutrino \cite{magnetic} and
the strength of right-handed interactions in the left-right
symmetric model\cite{left} among many others.  In this letter we
consider supernova SN1987A where the energy available for the
particles inside the star is about a few hundred of $MeV$ which is
far from the confirmed constraint on $\Lambda_{NC}$ to find new
bounds on the parameter of noncommutativity.  For this purpose one
should consider those new interactions which have not any
counterparts in the usual space.  In fact the new interactions which
appear in the theory as a $\theta$-correction to the existed
interactions inside the supernova, can not put a stringent bound on
the NC-parameter.  As a candidate we can consider the coupling of
right handed neutrino with photon in the NCSM which has not any
counterpart in the usual standard model.  To this end we briefly
introduce the direct coupling of the right handed neutrino with
photon in the NCSM.

There are two approaches to construct the standard model in the non
commutative space.  In the simplest one the gauge group is
$SU(3)_c\times SU(2)_L\times U(1)_Y$ in which the number of gauge
fields, couplings and particles are the same as the ordinary one
\cite{sm}.  In the second approach the gauge group is $U(3)\times
U(2)\times U(1)$ which is reduced to $SU(3)_c\times SU(2)_L\times
U(1)_Y$ by an appropriate symmetry breaking \cite{nogo-sm}. However,
in the both versions, new interactions with respect to the ordinary
standard model will appear due to the star product and the
Seiberg-Witten, SW, map.  Here we consider the former case and
denote an infinitesimal non-commutative local gauge transformation
of the fields contents of the theory as $
 \delta \widehat{\Psi}=i\rho_\Psi(\Lambda) \star \widehat{\Psi}$,
  where $\Lambda$ is a gauge parameter and $\rho_\Psi(\Lambda)$ is a
representation which is  carried by the matter or Higgs fields.  The
fields with hat are the non-commutative fields which can be written
as a function of ordinary fields using appropriate SW-maps.  The
right handed neutrino, $\widehat{\nu}_R$, as a neutral-hyper-charged
particle in the
 standard model can be transformed as
 $\de\widehat{\nu}_R=i\widehat{\La}^\prime\star\widehat{\nu}_R-i\widehat{\nu}_R\star\widehat{\La}^\prime$ to preserve the gauge symmetry.
 In fact in a non-commutative setting the non-commutative gauge
 boson ${B}_\mu$, compatible with the non-commutative gauge
 transformation, couples to a neutral matter field $\widehat{\Psi}^0$
 as $\widehat{{\cal D}}_\mu \widehat{\Psi}^0=\p_\mu
\widehat{\Psi}^0-i\lb\widehat{{B}}_\mu,\widehat{\Psi}^0\rb$,
 with $\delta\widehat{\Psi}^0=i[\widehat{\Lambda}\stackrel{\star}{,}\widehat{\Psi}^0]$.
  In other words, to introduce the neutrino-photon interaction in the minimal NCSM,
one can define the adjoint representation in the covariant
derivative for the neutral particle as is already done in the NCQED.
In other words in the frame work of NCQED it was shown that the
neutral particles interact with photons if they transform under U(1)
in a similar way as in the adjoint representation of a non-Abelian
gauge theory \cite{pn}. The main difference in the NCSM is $U(1)_Y$
instead of $U(1)_{EM}$. Therefore, in a gauge invariant manner, only
neutral hyper charge particle in the standard model can couple to
the hyper gauge field. The only particle with zero hyper charge in
the SM is the right handed neutrino therefore the covariant
derivative for this particle, to the lowest order, can be written as
$\widehat{D}_{\mu}\widehat{\psi}_{\nu_R}={\partial}_\mu\widehat{\psi}_{\nu_R}+
e{\theta}^{\nu\rho}{\partial}_\nu\widehat{B}_\mu
{\partial}_\rho\widehat{\psi}_{\nu_R}$, \cite{hez}, in which
$\widehat{\psi}_{\nu_R}$ and $\widehat{B}$, respectively, denote the
NC-fields of the right handed neutrino and the hyper charge with
their own expansion up to the lowest order in the NC-space as $
\widehat{\psi}_{\nu_R}={\psi}_{\nu_R} +
e{\theta}^{\nu\rho}B_\rho{\partial}_\nu{\psi}_{\nu_R}$ and $
\widehat{B}_\mu=B_\mu + e{\theta}^{\nu\rho}B_\rho[{\partial}_\nu
B_\mu-\frac{1}{2}{\partial}_\mu B_\nu]$. Consequently, Lagrangian
density for the right handed neutrino part of NCSM can be written as
follows \cite{hez}
\begin{eqnarray}
&{\cal{L}}_{\nu_R}&=i{\bar{\psi}}\partial\!\!\!\!/\psi +
ie{\theta}^{\mu\nu}[\partial_\mu{\bar{\psi}}B_\nu\gamma^\rho({\partial}_\rho\psi)
\nonumber\\& &
-\partial_\rho{\bar{\psi}}B_\nu\gamma^\rho({\partial}_\mu\psi) +
{\bar{\psi}}( \partial_\mu B_\rho
)\gamma^\rho({\partial}_\nu\psi)]\label{Lagden},
\end{eqnarray}
where $B$ in terms of the photon and the $Z$-gauge boson fields is
$B= \cos\theta_W A-\sin\theta_W Z$. Therefore the Feynman rules for
$\gamma\nu\bar\nu$ and $Z\nu\bar\nu$ vertices can be obtained from
the Lagrangian (\ref{Lagden}) as:
\begin{equation}
\Gamma^\mu_{\gamma\nu\bar\nu}=i{\frac
{e}{2}}cos\theta_W(1+\gamma_5)(\theta^{\mu\nu}k_\nu q \!\!\!/ +
\theta^{\rho\mu}q_\rho k \!\!\!/ + \theta^{\nu\rho}k_\nu
q_\rho\gamma^\mu),\label{gnn}
\end{equation}
and
 \begin{equation}
\Gamma^\mu_{Z\nu\bar\nu}=-i{\frac
{e}{2}}sin\theta_W(1+\gamma_5)(\theta^{\mu\nu}k_\nu q\!\!\!/ +
\theta^{\rho\mu}q_\rho k\!\!\!/ + \theta^{\nu\rho}k_\nu
q_\rho\gamma^\mu).\label{znn}
\end{equation}
  Since
the other particles in the SM, even the left handed neutrino, all
have nonzero hyper charge, the remaining parts of the SM in the
noncommutative space do not change.  It should be noted that for
neutrino as a neutral particle in the NCQED, as well as QED, in
contrast with the standard model, there is no constraint on the mass
or even the chirality of the neutrino. In the standard model,
neutrino is massless and only the left handed one has weak
interaction while the right handed neutrino, if existing, has an
expectator role in all reactions.

Now we ready to calculate the impact of this new interaction on the
luminosity of the supernova 1987A.  Inside a supernova, several
particle and nucleon processes contribute to neutrino production,
absorption and neutrino scattering.  In fact the neutrino
interactions are essential in producing and powering the supernova
explosion.  The neutrino spectrum is always characterized by a short
peak of electron
 neutrinos generated during the neutronization phase carrying 1$\%$ of the total
 energy and a second release of neutrinos of all flavors stemming from thermal pair processes
(pair annihilation, plasmon decay, photoneutrino and Bremsstrahlung)
releasing 99$\%$ of the energy.  Meanwhile since the nucleon density
is huge, the dominant lowest-order process for the emission of new
particles in a supernova core tends to be nucleon bremsstrahlung.
However in the calculation of nucleon-nucleon interactions, one
encounters the nonperturbative part of the strong interactions which
is not well known.  Furthermore, if the interaction of new particles
with all nucleons is the same, dipole emission in the nucleon
bremsstrahlung is suppressed in the nonrelativistic limit.
Therefore, one expects that the smaller leptonic density be
compensated by the fact that the pair annihilation channel is not
suppressed \cite{pair}-\cite{pair-unpar}.  Consequently pair
annihilation can be considered as a competent source of new
particles producing in the core of supernova.  In the absence of the
left-right symmetric model for the electroweak interaction, the
right handed neutrinos can be produced in the supernova via pair
annihilation in the noncommutative space
\begin{equation}
f+{\bar f}\rightarrow\nu_R+{\bar\nu_R}.\label{pair}
\end{equation}
As one can easily see, although at the lowest order in the NC-space,
the annihilation can be proceeded via the one photon exchange as
well as the one Z-boson exchange, the latter one is highly
suppressed because of the Z-boson propagator which leads to the
additional power of the Fermi coupling constant. Therefore we only
consider the one photon exchange diagram at the tree level and its
crossed amplitude for charged fermions specially for the electrons
and muons. Indeed the temperature of the proto-neutron star is
sufficient to generate a muonic number density as large as the
number density of the massless fermions.  Once the right handed
neutrinos are produced, they have mean-free-paths in the supernova
core which are determined via the cross sections for the scattering
processes as follows
\begin{eqnarray}
&&\nu_R +f\rightarrow \nu_R+f,\nonumber\\ &&\nu_R+p\rightarrow
\nu_R+p.\label{scatt}
\end{eqnarray}
In light of this situation, here we consider the observed neutrinos
from SN 1987A \cite{exp} which is in agreement with the theoretical
picture for the Type-II supernovae.   The impact of the new
interaction on the data of the SN 1987A is twofold:
\begin{enumerate}
\item  If the strength of the new coupling which depends inversely on the
parameter of noncommutativity, $\Lambda_{NC}$, leads to a mean free
path larger than the core radius, the right handed neutrinos can
escape the proto-neutron star.  This provides a new channel for
energy loss which increases with decreasing $\Lambda_{NC}$.
Therefore the comparison of the $\nu_R$-luminosity with the well
known data from the SN 1987A can put a lower bound on
$\Lambda_{NC}$.
\item Meanwhile the absorbtion cross section increases as $\Lambda_{NC}$ decreases and therefore for a certain range
 of $\Lambda_{NC}$ the neutron star becomes optically thick for the
 right handed neutrinos. In this case they have mean free path smaller than the core
radius and can be trapped inside the $\nu_R$-sphere.  In this
situation the luminosity can be approximated by black body radiation
which in turn decreases as $\Lambda_{NC}$ decreases.  In other
words, there is an upper bound on the parameter of noncommutativity
in this case.
\end{enumerate}
{\bf Free Streaming Case:} In the NC-space there is no constraint on
mass of the right handed neutrino. Therefore in the NC-space, the
light right handed neutrinos can be produced in the SN via the
fermion pair annihilation given in Eq.(\ref{pair}).  By using the
$\gamma\nu_R\nu_R$-vertex given in Eq.(\ref{gnn}) and neglecting the
mass of neutrino, the spin-averaged amplitude in terms of the
Mandelstam variables can be obtained as
\begin{equation}
\overline{\mid {\cal M}\mid}^2=e^4cos^4\th_W[k\cdot\th\cdot
p]^2{\frac{t^2+u^2+m^2_fs}{s^2}},
\end{equation}
where $k$ and $p$ are the momenta of $f$ and $\nu_R$, respectively.
Here for simplicity we calculate the total cross section in the
center of mass frame.  To this end we define
\begin{eqnarray}
&&{\vec\theta_s}=(\theta_{23}, \theta_{31},
\theta_{12})=(\theta_{s\parallel}, \theta_{s\perp}),\nonumber\\
&&{\vec\theta_t}=(\theta_{01}, \theta_{02},
\theta_{03})=(\theta_{t\parallel}, \theta_{t\perp}),
\end{eqnarray}
where $\perp$ and $\parallel$ mean perpendicular to and parallel
with the surface contains $\vec p$ and $\vec k$.  For the electron
pair annihilation, we can neglect the mass of electron with respect
to the center of mass energy, and after averaging on the angle of
$\vec \theta$ with the $\vec p$ and $\vec k$, to find
\begin{equation}
\sigma(ee^+\rightarrow\nu_R\bar\nu_R)\simeq
{\frac{e^4cos^4\th_W}{30\pi}}\Theta^2E^2=0.01(\frac{M_W}{\Lambda_{NC}})^4G_F^2E^2\label{sigma-pair},
\end{equation}
where $\Theta=[
5\mid\theta_{t\parallel}\mid^2/8+3\mid\theta_{s\perp}\mid^2/8]$ and
$\Lambda_{NC}=1/\sqrt\Theta$.

If the volume energy-loss rate of the SN by any new mechanism
becomes greater than $Q\sim 3 \times 10^{33}\,\,\, ergs\,\,\,
cm^{-3}\,\,\, s^{-1} $ then it will remove sufficient energy from
the explosion to invalidate the current understanding of the
SN1987A-neutrino signal \cite{sn}.  Meanwhile the new mechanism in
the NC-space, Eq.(\ref{pair}), leads to the energy emitted by the
supernova per unit time and unit volume as follows
\begin{equation}
Q=n(e^+)n(e^-)\langle\sigma(ee^+\rightarrow\nu_R\bar\nu_R)v_{rel}(E_\nu+E_{\bar\nu})\rangle\label{Q-pair},
\end{equation}
where $n(e^\pm)$ is the number density of electron (positron) and
$v_{rel}$ is the relative velocity of $e^-$ and $e^+$.  Therefore
Eq.(\ref{sigma-pair}) and Eq.(\ref{Q-pair}) for zero chemical
potential and equal number density of electron and neutron result in
\begin{equation}
Q=2.5\times10^{-3}(\frac{M_W}{\Lambda_{NC}})^4G_F^2{\frac{16}{3\pi^4}}T^9\int_0^\infty{\frac{x^3}{e^x+1}}dx\int_0^\infty{\frac{x^4}{e^x+1}}dx,
\end{equation}
where for $T=50$ one finds $Q\sim
(\frac{M_W}{\Lambda_{NC}})^4\times10^{40}\,\,\, ergs\,\,\,
cm^{-3}\,\,\, s^{-1}$ or $\Lambda_{NC}\gtrsim 4.5 $ TeV.  The
presence of a chemical potential reduces the number of pairs but for
the muon with the rest mass of 106 MeV ignoring the chemical
potential does not lead to a substantial error.  However for the
pairs of muons, there is a suppression factor with respect to the
electrons of about 0.5 comeing from the relative velocity of muons.
Therefore for $T=50$ one finds $\Lambda_{NC}\gtrsim 3.7 $ TeV.

{\bf Trapping Case:} If the interaction of the right handed neutrino
in the NC-space is strong enough it can be trapped and its
subsequent thermalization can reduce the luminosity.  The most
relevant reactions which contribute to trapping are given in
Eq.(\ref{scatt}).  In order to find the luminosity in this case we
need to estimate the radius of the $\nu_R$-sphere.  Now for
simplicity  we neglect the mass of the charged fermion and restrict
ourselves to the space noncommutativity (i.e. $\theta_t=0$), to
obtain the differential cross section as follows
\begin{equation}
{\frac{d\sigma}{d\Omega}}={\frac{e^4cos^4\th_W}{256\pi^2}}E^2\mid\theta_{s\perp}\mid^2{\frac{(1+cos\alpha)(4+(1+cos\alpha)^2)}{1-cos\alpha}},
\end{equation}
where $\alpha$ is the scattering angle.  It should be noted that the
differential cross section as usual has a singularity at
$cos\alpha=1$ but here the singularity is only logarithmic.  To
obtain the total cross section we need a cutoff to regularize the
divergence when $\alpha$ goes to zero.  In fact neutrinos are
confined in the $\nu_R$-sphere with a radius of the order of the
core radius, $R_C$, that constrains their transverse momentum to be
at least $p_{\perp}\thicksim \frac{1}{R_C}$ or the scattering angle
should be greater than $\alpha=\frac{p_{\perp}}{E}\thicksim
10^{-19}$. Alternatively, the screening effect in the core can
provide a mass of the order of $\mu=V^{-\frac{1}{3}}$ to the photon
which in turn changes  the variable $t\rightarrow t-\mu^2$ or
$1-cos\alpha$ to $1-cos\alpha + \frac{\mu^2}{2E^2} $ which is
regular at $cos\alpha=1$ and is equivalent to $\alpha\gtrsim
10^{-19}$. Therefore the total cross section can be easily obtained
as
\begin{equation}
\sigma=3.28(\frac{M_W}{\Lambda_{NC}})^4G_F^2E^2\label{trap-cross}.
\end{equation}
Eq.(\ref{trap-cross}) can be used to calculate the mean free path of
the right handed neutrino.  For constant temperature and core number
density one can easily find the right handed neutrino mean free path
as
\begin{equation}
l_{\nu_R}=\frac{1}{n\sigma}=\{95.12(\frac{M_W}{\Lambda_{NC}})^4
(\frac{n}{n_C})(\frac{T}{50})^2\}^{-1}\,\,\,\,\,m\label{meanfree},
\end{equation}
where $E^2\thicksim 10T^2$ and $n_C=2.4\times10^{44}\,\,\,m^{-3}$.
In the trapping case $l_{\nu_R}\lesssim R_C$ which for $R_C\thicksim
10^4\,\,\,m$ and $T=50 \,\,\,MeV$ leads to $\Lambda_{NC}\lesssim 31
M_W$.  If one considers the variation of temperature and number
density in the core a more realistic bound on the $\Lambda_{NC}$ can
be found.  For instance, if $n(R)=n_C(\frac{R}{R_C})^m$ and
$T(R)=T_C(\frac{R_C}{R})^{\frac{m}{3}}$, for an appropriate value of
$m$ see \cite{sterile} and \cite{left}, one finds from
Eq.(\ref{trap-cross}) $\Lambda_{NC}\lesssim 1.1\,\,\,TeV$.

Now let us summarized our results. In this letter we showed that the
direct coupling of the right handed neutrino with photon in the NCSM
can provide a new channel for the energy loss of SN 1987A.  Since
NC-calculations are perturbative and are correct only up to the
energy scale of the NC-system $\lesssim\Lambda_{NC}$, therefore SN
1987A as a system with energy scale of a few hundred of MeV is an
excellent candidate to find bounds on $\Lambda_{NC}$.  We found for
the weak coupling that $\Lambda_{NC}\gtrsim 3.7 TeV$ which is more
stringent than the
 other bounds obtained from LEP and LHC considerations.  Meanwhile
 for the strong coupling, the trapping case, we obtained $\Lambda_{NC}\lesssim 1
 TeV$.  In fact constraints on the energy loss of SN 1987A exclude
 the range $1 TeV\lesssim\Lambda_{NC}\lesssim 3.7 TeV$ for the NC-parameter.
The lower bound can be considered seriously  because
 the obtained bounds from LEP and LHC considerations are usually
 based on the fact that $\Lambda_{NC}\gtrsim $ the energy scale of
 the system which may not be correct.  Therefore $\Lambda_{NC}\lesssim 1
 TeV$ is consistent with the existing bounds on $\Lambda_{NC}$ and raises our hopes to find the NC-effects in the LHC or even in the
 LEP.  Finally it should be noted that  the well
 known processes which can successfully describe
 the energy loss of SN 1987A  in the ordinary space also receive the
 NC-corrections.  But theses corrections to the
usual vertices are too small to change the energy loss of the
supernova. For instance, in the production of the left handed
neutrino in the ordinary space the cross section is of the order of
$\sigma_0\sim G^2 E^2$ and at the lowest order in the NC-space its
correction is of the order of $\sigma_{NC}\sim \theta G^2 E^4$
where for $\Lambda=\frac{1}{\sqrt\theta}\sim 1 TeV$ and for the
energy range $\sim 10 MeV$ the fraction $
{\frac{\sigma_{NC}}{\sigma_0}}\sim \theta E^2\sim 10^{-10}$
 is negligible.  Therefore the energy loss of the supernova via
 the usual channels do not change in the NC-space.


\end{document}